\documentclass[conference]{IEEEtran}
\IEEEoverridecommandlockouts

\usepackage{amsmath,amssymb,amsfonts}
\usepackage{algorithmic}
\usepackage{graphicx}
\usepackage{textcomp}
\usepackage{xcolor}
\def\BibTeX{{\rm B\kern-.05em{\sc i\kern-.025em b}\kern-.08em
    T\kern-.1667em\lower.7ex\hbox{E}\kern-.125emX}}

\usepackage{makecell}
\usepackage{booktabs}
\usepackage{multirow}
\usepackage{subcaption}
\usepackage{tabularx,array}
\usepackage{graphicx} 
\usepackage{mathtools}
\usepackage{bm}      
\usepackage[acronym,nonumberlist,toc,section=section]{glossaries}
\usepackage{hyperref}
\usepackage{setspace}
\usepackage{cuted}
\usepackage{stfloats}

\usepackage{optidef}
\usepackage{etoolbox}

\hypersetup{
    colorlinks,
    linkcolor={red!50!black},
    citecolor={blue!50!black},
    urlcolor={blue!80!black}
}

\newcounter{assumption}
\usepackage{lipsum}
\usepackage{xcolor}

\usepackage{calc}

\newcommand{\xboxed}[2][\fboxsep]{%
  \begingroup
  \setlength{\fboxsep}{#1}%
  \boxed{#2}%
  \endgroup
}

\usepackage{nomencl}
\makenomenclature

\makeatletter
\renewcommand\@makefntext[1]{%
  \noindent\@makefnmark\ #1%
}
\makeatother

\begin{document}

\makeatletter
\newcommand{\linebreakand}{%
  \end{@IEEEauthorhalign}
  \hfill\mbox{}\par
  \mbox{}\hfill\begin{@IEEEauthorhalign}
}
\makeatother

\title{Peak-Load Pricing and Investment Cost Recovery with Duration-Limited Storage\\
\thanks{This material is based upon work supported by the MIT Energy Initiative's
Future Energy Systems Center.}}

\author{\IEEEauthorblockN{Daniel Shen}
\IEEEauthorblockA{\textit{Department of Electrical Engineering \& Computer Science} \\
\textit{Massachusetts Institute of Technology}\\
Cambridge, USA \\
oski@mit.edu}
\and
\IEEEauthorblockN{Marija Ilic}
\IEEEauthorblockA{
\textit{Department of Electrical Engineering \& Computer Science} \\
\textit{Massachusetts Institute of Technology}\\
Cambridge, USA \\
ilic@mit.edu}
\linebreakand 
\IEEEauthorblockN{John Parsons}
\IEEEauthorblockA{
\textit{Center for Energy and Environmental Policy Research} \\
\textit{Massachusetts Institute of Technology}\\
Cambridge, USA \\
jparsons@mit.edu}
}

\maketitle

\begin{abstract}
Energy storage shifts energy from off-peak periods to on-peak periods. Unlike
conventional generation, storage is duration-limited: the stored energy
capacity constrains the duration over which it can supply power. To understand
how these constraints affect optimal pricing and investment decisions, we
extend the classic two-period peak-load pricing model to include
duration-limited storage. By adopting assumptions typical of solar-dominated
systems, we link on- and off-peak prices to storage investment costs,
round-trip efficiency, and the duration of the peak period. The bulk of the
scarcity premium from on-peak prices is associated with the fixed costs of
storage as opposed to variable costs stemming from round-trip efficiency
losses. Unlike conventional generators, the binding duration constraints lead
storage to recover energy capacity costs on a per-peak-event basis instead of
amortizing these costs over total peak hours. A numerical example illustrates
the implications for equilibrium prices and capacity investment.
\end{abstract}

\begin{IEEEkeywords}
electricity markets, energy storage, peak-load pricing, scarcity pricing
\end{IEEEkeywords}

%
\section{Introduction}
\label{sec:intro}

The installation of grid-scale energy storage is accelerating.
In the United States, storage comprised 29\% of expected new generation
capacity in 2025, up from 23\% in 2024 and 17\% in 2023
\cite{u.s.energyinformationadministrationSolarBatteryStorage2025}.
Unlike conventional generators, storage does not generate electricity directly,
but rather shifts an amount of electricity from off-peak hours of low net load
to on-peak hours of high net load. This arbitrage reduces the utilization of
more expensive generators and decreases system costs.

These distinct characteristics and accompanying benefits have motivated
consideration of how storage can be further integrated into electricity
markets. For instance, in the United States, FERC Order 841 (passed in 2018)
directed market operators to remove barriers to the participation of battery
storage in wholesale markets, with the goal of enabling storage to compete
alongside conventional generators. This growth of grid-scale energy storage
prompts the following questions:

\begin{itemize}
    \item \textit{What market prices will incentivize efficient investment in
          generation and storage?}
    \item \textit{How are such prices linked to the characteristics of the storage
          technology?}
\end{itemize}

The electricity literature empirically addresses these questions. For instance,
prior work establishes that storage reduces the price spread between high and
low net demand periods \cite{sioshansiEstimatingValueElectricity2009,
zamani-dehkordiPriceImpactAssessment2017, lampLargescaleBatteryStorage2022}.
We build upon these results from an alternative conceptual lens, aiming for a
more foundational understanding of storage in power systems. For example, a
canonical benchmark is that in equilibrium, peakers are sized to just break
even on their fixed costs during scarcity hours. What would the corresponding
principles be for systems with storage?

Our lens aligns with an analytically-focused strand of the literature
where storage shapes equilibrium outcomes.
Reference \cite{antweilerNewMeritOrder2025} examines the equilibrium prices in
a system consisting solely of variable renewable energy (VRE), storage, and
demand response. Their ``new merit order'' has steps in the price schedule
corresponding to efficiencies of different storage technologies. In contrast,
\cite{korpasOptimalityConditionsCost2020} utilizes a duration-curve
incorporating VRE, storage, and conventional generation, and shows that
storage-induced prices support capacity investments in equilibrium. However,
these works critically make the assumption that storage is power-limited but
duration-\textit{unlimited}, i.e. the stored energy constraint is nonbinding.

This duration limit is a defining characteristic of storage, since a unit that
depletes its stored energy during a peak period cannot continue to provide
supply. To preserve this characteristic while gleaning
analytical insights, we extend the Steiner/Boiteux peak-load pricing model
\cite{steinerPeakLoadsEfficient1957, marcelboiteuxPeakLoadPricing1960}.
Although simple, this peak-load model captures the fundamental dynamics
affecting resource investments and prices. Prior extensions of such models to
incorporate storage make a duration-unlimited assumption
\cite{nguyenProblemsPeakLoads1976} or duration-limited but power-unlimited
assumption regarding the storage resource
\cite{schmalenseeCompetitiveEnergyStorage2022}. In contrast, we impose limits
on both the power \textit{and} stored energy capacity to analyze the scarcity
premium associated with these constraints. By applying several simplifying
assumptions which are reasonable in a system where the net load peak is
periodic, predictable, and accompanied by long off-peak periods, we relate
electricity prices in equilibrium to storage parameters. \footnote{Our
assumptions mirror that of Schmalensee in
\cite{schmalenseeCompetitiveEnergyStorage2022}, who couches his system setup in
a solar-dominated grid.} In particular, we find:

\begin{itemize}
    \item Scarcity prices are primarily driven by storage fixed costs, rather
          than operating costs from round-trip losses.
    \item Duration limits introduce a fixed cost component to the on-peak price
          which contributes per peak event instead of per total peak duration,
          so scarcity events cannot be fully aggregated into
          load durations to evaluate storage.
    \item If the storage is optimally sized and operated, the resulting prices
          guarantee investment cost recovery.
\end{itemize}

The paper proceeds as follows: Section \ref{sec:model-intro} presents
our two-period peak-load model. Section \ref{sec:energy-price-derivations} derives
analytic expressions for the price schedule and demonstrates
the points above. Section \ref{sec:example-model} presents a numerical
example to illustrate how storage reshapes the capacity mix and price
structure. Finally, Section \ref{sec:conclusion} concludes.
\section{Model}
\label{sec:model-intro}
\subsection{Setup}

\begin{figure*}[t]
\centering
    \input{00-formulation}
\vspace{-0.5cm}
\end{figure*}

In the classic peak-load model, a social planner jointly optimizes
investment and operation of conventional generators to maximize total surplus
net of investment costs. We add a duration-limited storage resource to this basic
peak-load model; our formulation is given by \eqref{eq:system-optimization}. We
consider two periods indexed by $i$, with duration $T_i$. Electricity demand
per period follows a downward-sloping inverse demand function $p_i$. The
decision variables are investments in $K$ (power capacity) and $E$ (stored
energy capacity) along with the operating decisions $q$ (per-period injections)
and $\ell$ (per-period consumptions). Dual variables associated with each
constraint are denoted in square brackets.

Baseload and peaker generators have fixed investment costs $I_B$ and $I_P$ per
unit of power capacity; these investment costs are normalized by the number of
peak cycles $n$ per year. Each technology $g \in \{B, P\}$ can produce up to
its installed capacity $K_g$ at a constant operating cost $c_g$.

In each period, supply and demand must balance per \eqref{const:powerbalance}.
The dual of \eqref{const:powerbalance} is the energy price $\lambda_i$ in
each period. Energy can be supplied by either the baseload, peaker, or storage
unit, while energy can be demanded by consumers or by storage for charging. The
storage unit can charge or discharge at a constant rate $q_i^+$ or $q_i^-$ in
each period, as long as this power is below the unit's power capacity $K_S$.
Any energy charged is subject to a round-trip efficiency $\eta$ per
\eqref{const:rte}. Nontrivial storage operation entails charging in one period
and discharging in the other. Thus, the aggregate energy balance
\eqref{const:rte} across both periods suffices to characterize storage
operation in our setting. Equations \eqref{const:energy-charge} and
\eqref{const:energy-discharge}
impose a stored energy limit $E$.

\subsection{Optimality Conditions}

The Lagrangian associated with (\ref{eq:system-optimization}) yields the
following subset of KKT conditions which are relevant to
our results:

\newcounter{kkt}
\newcommand{\kkttag}{\refstepcounter{kkt}\tag{KKT-\thekkt}}

\newcommand{\stationarity}[1]{\frac{\partial\mathcal{L}}{\partial#1}}

\begingroup
\allowdisplaybreaks

\begin{align}
    \stationarity{q_i^+} &= 0 = -\lambda_i T_i - \sigma_i^+ T_i +
    \zeta_i^+ T_i + \mu \eta T_i - \gamma^+ T_i \eta
    \kkttag \label{eq:stationary-charge} \\
    \stationarity{q_i^-} &= 0 = \lambda_i T_i - \sigma_i^- T_i + \zeta_i^- T_i - \mu T_i - \gamma^- T_i
    \kkttag \label{eq:stationary-discharge} \\
    \stationarity{K_s} &= 0 = -\frac{1}{n} I_{s,q} + \sum_i T_i (\sigma_i^+ + \sigma_i^-)
    \kkttag \label{eq:stationary-storagek} \\
    \stationarity{E} &= 0 = - \frac{1}{n} I_{s,E} + \gamma^+ + \gamma^-
    \kkttag \label{eq:stationary-storagee}
\end{align}

\endgroup
\vspace{-0.25cm}

\section{Optimal Prices}
\label{sec:energy-price-derivations}

\subsection{On-Peak Prices}
\label{sec:on-peak-prices}

We now characterize the relationship between on- and off-peak prices. To obtain
closed-form expressions relating these prices to the parameters of
the storage technology, we impose the following three simplifying assumptions,
with their associated implications indicated by the $\implies$ symbol:

\begin{list}{}%
  {%
    \leftmargin=5pt
    \itemindent=0pt
    \labelsep=0pt
    \labelwidth=0pt
    \itemsep=\parskip
    \parsep=0pt
    \topsep=0pt
  }
\item
\refstepcounter{assumption}%
\vspace{0.1cm}
\textbf{Assumption 1: Price ordering.}
The optimal consumption $\ell$ for each period creates a difference
between on-peak and off-peak prices sufficient to induce storage
investment.
\footnote{Two alternatives are possible here. First, if the price difference
is below a threshold, storage will not be economical and the problem reduces to
a standard peak-load model without storage. Second, if the off-peak price is
the higher price due to the shape of the demand functions, we can simply
relabel the period with the higher price as the ``on-peak'' period.}
\newline
$\implies$ There is a nonzero amount of storage capacity investment;
storage charges off-peak and discharges on-peak.
\label{assume:cycling}
\vspace{0.15cm}
\item
\refstepcounter{assumption}%
\textbf{Assumption 2: Sufficient off-peak duration.}
The off-peak period satisfies
$T_{\text{onpeak}} < \eta T_{\text{offpeak}}
\implies$ The storage power constraint binds only during on-peak discharge.
\label{assume:off-peak-binding}
\vspace{0.15cm}
\item
\refstepcounter{assumption}%
\textbf{Assumption 3: No inter-period carryover value.}
Stored energy has no continuation value during the off-peak period.
$\implies$ Stored energy is zero at the end of the on-peak period.
\label{assume:no-carryover}
\vspace{-0.2cm}
\end{list}
These assumptions can reflect systems with a single dominant daily peak and
long periods of low net demand, as may be typical of solar-heavy systems
where storage charges midday and discharges to meet the evening peak.

Combining conditions (\ref{eq:stationary-charge}) and
(\ref{eq:stationary-discharge}) gives:

\begin{align}
    \mu &= \frac{\lambda_i + \sigma_i^+ + \gamma^+\eta - \zeta^+_i}{\eta} \\
    &= \lambda_j - \sigma_j^- - \gamma^- - \zeta_j^-
    \quad \forall i,j \in \{ \text{on-peak, off-peak} \} \notag
\end{align}

Substituting $i$ = off-peak and $j$ = on-peak:

\begin{equation}
    \lambda_{offp} + \sigma_{offp}^+ + \gamma^+ \eta - \zeta^+_{offp} =
    \eta (\lambda_{onp} - \sigma_{onp}^- - \gamma^- -  \zeta_{onp}^-)
\end{equation}

Assumption (\ref{assume:cycling}) implies constraints
\eqref{const:chargemin} and \eqref{const:dischargemin}
are not binding and the corresponding duals
$\zeta^+_{offp}$ and $\zeta^-_{onp}$
are zero. Similarly, Assumption (\ref{assume:off-peak-binding}) yields
$\sigma_{offp}^+ = 0$. Substituting
$\zeta^+_{offp}=\zeta^-_{onp}=\sigma_{offp}^+=0$ and rearranging:

\begin{equation}
    \lambda_{offp} = \eta \lambda_{onp} - \eta
    \left( \sigma^-_{onp} + \gamma^+ + \gamma^- \right)
    \label{eq:penultimate-energy-price}
\end{equation}

Assumptions
\eqref{assume:off-peak-binding} and \eqref{assume:no-carryover} taken together
imply that of the storage power capacity constraints
\eqref{const:chargemax}, \eqref{const:dischargemax}, only the on-peak
discharging constraint is binding:
\begin{equation}
    \label{eq:sigmas-zero}
    \sigma^-_{onp} > 0, \sigma^+_{onp} = \sigma^+_{offp} = \sigma^-_{offp} = 0
\end{equation}

Combining \eqref{eq:sigmas-zero} with
\eqref{eq:stationary-storagek} and
\eqref{eq:stationary-storagee} allows us to eliminate all dual
variables from \eqref{eq:penultimate-energy-price} which are not
associated with the power balance constraint \eqref{const:powerbalance}.
The final relationship between energy prices in the on-peak and off-peak
periods under duration-limited storage is thus:

\begin{equation}
    \xboxed[20\fboxrule]{
        \label{eq:on-peak-off-peak-prices}
        \lambda_{onp} = \frac{\lambda_{offp}}{\eta} +
        \frac{1}{n}\left( I_{s, E} + \frac{I_{s,q}}{{T_{onp}}}\right)
    }
\end{equation}

Equation \eqref{eq:on-peak-off-peak-prices} decomposes the on-peak price into a
variable-cost component and a fixed-cost component. The first component
$\lambda_{offp} / \eta$ reflects the costs of purchasing energy for use
later and is analogous to the variable cost of a conventional generator. The
second component reflects the additional premium required to cover the fixed
investment costs of the storage system. This mirrors the cost-recovery
condition for a peaker plant in the canonical two-period model:

\begin{equation}
    \lambda_{onp} = c_P + \frac{1}{n}\left(\frac{I_P}{T_{onp}}\right)
    \label{eq:peaker-price}
\end{equation}
but with \eqref{eq:on-peak-off-peak-prices} including an additional term $I_{s,E} /
n$ that reflects the fixed costs of stored energy capacity. Notably, this energy
capacity term is not scaled by the on-peak duration $T_{onp}$; the energy
capacity premium is independent of the total duration of on-peak hours
across a year ($n T_{onp}$) and is solely dependent on the \textit{number} of
peaks $n$ during a year. This differs from \eqref{eq:peaker-price} where the
resource's fixed costs are amortized over the total duration of peaks.

In practice, the price differential in \eqref{eq:on-peak-off-peak-prices} is
dominated by the fixed costs rather than efficiency losses. For lithium-ion
batteries, the NREL Annual Technology Baseline reports annualized fixed costs
of approximately $I_{s,q}=\$36$k/MW-year and $I_{s,E}=\$31$k/MWh-year. Assuming
an off-peak price of \$20/MWh, a round-trip efficiency of $\eta=85\%$, and an
on-peak duration of $T_{onp}=4$ hours, the efficiency-related term contributes
a price premium of \$23.5/MWh, while the fixed-cost term contributes
\$109.6/MWh. Fixed costs account for nearly five times the price impact of
efficiency losses --- equilibrium on-peak prices must rise well above variable
operating costs to provide adequate incentives for storage investment.
\footnote{Constraint
\eqref{const:powerbalance} permits prices of either sign. Our results thus
depend only on the magnitude of price \textit{spread} and not on the sign of
price levels. When both prices are negative, storage is remunerated by off-peak
charging, and the on-peak period acts as a binding rate constraint on resetting
the energy state.
}

\subsection{Cost Recovery}

Next, we show that setting energy prices in the ``standard''
manner -- the dual $\lambda$ of the power balance constraint
(\ref{const:powerbalance}) -- guarantees cost recovery for the storage
system if the three assumptions in Section \ref{sec:on-peak-prices} hold.

The total storage investment costs are $I_{S, q} K_S + I_{S, E} E$.
The operating profits are the difference between revenues from energy sold during
on-peak periods and the cost of energy purchased during off-peak periods.
Under dual-based pricing using (\ref{const:powerbalance}), the break-even
condition equating investment costs to operating profits over $n$ cycles is:

\begin{multline}
    \label{eq:breakeven-condition}
    I_{S, q} K_S + I_{S, E} E =\\
        n \left(\lambda_{onp} T_{onp} q^-_{onp} -
        \lambda_{offp} T_{offp} q^+_{offp} \right)
\end{multline}

The efficiency constraint \eqref{const:rte} and
Assumption (\ref{assume:no-carryover}) imply the
$T_{onp} q^-_{onp}$ units of energy sold on-peak must be matched by
$\frac{1}{\eta} T_{onp} q^-_{onp}$ units of energy purchased during the
off-peak period, or:

\begin{equation}
    \label{eq:purchase-peak}
    T_{offp} q^+_{offp} = \frac{1}{\eta} T_{onp} q^-_{onp}
\end{equation}

The optimal action for storage is to discharge at full
capacity $K_S$ during the on-peak period. If this were not
the case, there would be uneconomical slack in the power capacity. Thus,

\begin{equation}
    \label{eq:onpeak-noslack}
    q^-_{onp} = K_S
\end{equation}

Combining (\ref{eq:on-peak-off-peak-prices}), (\ref{eq:breakeven-condition}),
(\ref{eq:purchase-peak}), and (\ref{eq:onpeak-noslack})
allows us to complete the proof of cost recovery:

\begin{equation}
    \label{eq:proof-of-recovery}
    \begin{gathered}
        n\left( \lambda_{onp}T_{onp} q^-_{onp}-
            \lambda_{offp}T_{offp} q^+_{offp} \right) \\
        = n K_S \left( \lambda_{onp}T_{onp} -
            \frac{1}{\eta}\lambda_{offp}T_{onp} \right) \\
        = n K_S T_{onp} \left[
            \left(
                \frac{\lambda_{offp}}{\eta} + \frac{1}{n}
                \left(
                    I_{S,E} + \frac{I_{S,q}}{T_{onp}}
                \right)
            \right)
            - \frac{1}{\eta}\lambda_{offp}
         \right] \\
    = K_S T_{onp}I_{S,E} + K_S I_{S, q} \\
    = E I_{S, E} + K_S I_{S, q}
    \end{gathered}
\end{equation}

Equation (\ref{eq:proof-of-recovery}) verifies the break-even condition in
(\ref{eq:breakeven-condition}) holds exactly. The final step equates $K_S
T_{onp}$ to $E$ because the optimal energy capacity $E$ is sized to support
discharging at the full power capacity $K_S$ throughout the entire on-peak
period. Thus, uniform pricing set by the energy balance dual
\eqref{const:powerbalance} in each period ensures an optimally sized storage
unit will recover its investment costs through energy profits alone.

\begin{table*}[tb!]
\centering
\small
\renewcommand{\arraystretch}{1.2}
\caption{Technology Cost Parameters (Stylized from NREL Annual Technology Baseline Values)}
\label{table:model-setup}

\begin{tabular}{m{0.13\textwidth} m{0.15\textwidth} m{0.2\textwidth} m{0.2\textwidth} m{0.2\textwidth}}
\hline
\textbf{Technology} & \textbf{Operating Cost} & \multicolumn{2}{c}{\textbf{Annualized Investment Cost}} & \textbf{Round-Trip Efficiency} \\ \cline{3-4}
                    & \textbf{(\$/MWh)}                & \textbf{Power (\$/MW-year)}    & \textbf{Energy (\$/MWh-year)}    & \textbf{(-)}                            \\ \hline
Peaker              & 100                     & 120,000                   & --                          & --                             \\
Baseload            & 20                      & 240,000                   & --                          & --                             \\
Storage             & --                      & 36,000                    & 31,000                      & 0.85                           \\ \hline
\end{tabular}
\end{table*}

\section{Example}
\label{sec:example-model}
We illustrate the results of the preceding section with a stylized numerical
example that highlights how storage alters equilibrium prices and
capacity investments. We construct two scenarios: one
``with storage'' and conventional generation and a counterfactual ``without
storage'' that only has conventional generation. The model is constructed in
Pyomo \cite{hartPyomoModelingSolving2011} and solved with Gurobi v12.0.3.

The system operation is treated as a representative day repeated $n$ = 365 times
to form an annual cycle.
Each day has an on-peak period and an off-peak period of a duration of four and
twenty hours, respectively. The demand in each period follows a linear inverse
demand function $p_i(z_i) = a_i - b_i \ell_i$;
the demand elasticity is set to 0.1 at baseline points of 10 GW \&
\$20/MWh for the off-peak period and 15 GW \& \$100/MWh for the on-peak period.

\begin{figure}[tbp]
  \centering
    \includegraphics[width=\linewidth]{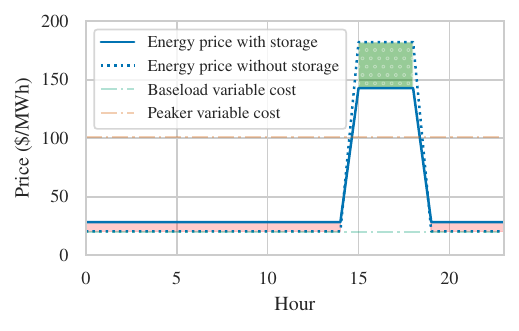}
    \caption{
      On- and off-peak energy prices with and without storage. The dotted green
      area highlights where storage decreases prices
      and the red area highlights where storage increases prices
      relative to the ``without storage'' scenario.
    }
  \label{fig:model-results-prices}
  \vspace{-0.5cm}
\end{figure}

Table \ref{table:model-setup} lists the technology characteristics used in the
example. The resulting prices and dispatch decisions in each period are
provided in Table \ref{table:operating-params}. The electricity prices
are also shown in Fig. \ref{fig:model-results-prices} for a 24 hour period.
Table \ref{table:generation-capacity} lists the generation capacity investments
in each scenario.

With storage, on-peak prices are \$182/MWh and exceed the variable cost
associated with charging prices and efficiency losses. Analogous to the classic
peak-load pricing result, this price level reflects the scarcity of
generation and enables investment recovery ---
\eqref{eq:on-peak-off-peak-prices} shows that the bulk of this scarcity premium
is driven by the storage's fixed costs, rather than the operating costs
associated with efficiency losses.

Our example utilizes parameters characteristic of lithium-ion storage. For
technologies with substantially lower round-trip efficiencies (e.g. hydrogen,
where efficiencies are 20 -- 40\%), the price premium associated with
round-trip losses would be higher. However, realistic scarcity prices
reflecting the value of lost load can easily exceed \$1000/MWh, dwarfing the
variable costs associated with efficiency losses even for low-efficiency
technologies. Thus, regardless of the storage technology, investment costs will
likely contribute to the bulk of on-peak equilibrium prices.

\begin{table*}[tb!]
\centering
\small
\renewcommand{\arraystretch}{0.95}
\caption{On-Peak and Off-Peak Prices and Dispatch Decisions}
\label{table:operating-params}

\begin{tabularx}{\textwidth}{@{}
    >{\raggedright\arraybackslash}p{0.15\textwidth}
    *{5}{>{\raggedright\arraybackslash}X}
    >{\raggedright\arraybackslash}p{0.02\textwidth}
    *{5}{>{\raggedright\arraybackslash}X}
@{}}
\toprule
\textbf{Scenario} &
\multicolumn{5}{c}{\textbf{On-Peak}} & &
\multicolumn{5}{c}{\textbf{Off-Peak}} \\
\cmidrule(lr){2-6} \cmidrule(lr){8-12}
& $\lambda$ (\$/MWh)
& $\ell$ \newline (GW)
& $q_P$ (GW)
& $q_B$ (GW)
& $q^-$ (GW)
&
& $\lambda$ (\$/MWh)
& $\ell$ \newline (GW)
& $q_P$ (GW)
& $q_B$ (GW)
& $q^+$ (GW) \\
\midrule
w/ storage    & 142 & 14.4 & 0   & 10.5 & 3.9 & & 28  & 9.6  & 0   & 10.5 & 0.9 \\
w/out storage & 182 & 13.8 & 3.8 & 10.0 & --  & & 20  & 10.0 & 0   & 10.0 & --  \\
\bottomrule
\end{tabularx}
\end{table*}

\begin{table}[tb!]
\centering
\small
\renewcommand{\arraystretch}{0.95}
\caption{Generation Capacity Investments (GW and GWh)}
\label{table:generation-capacity}

\begin{tabularx}{\columnwidth}{@{}
    >{\raggedright\arraybackslash}p{0.28\columnwidth}
    *{4}{>{\raggedright\arraybackslash}X}
@{}}
\toprule
\textbf{Scenario} &
$K_B$  &
$K_P$  &
$K_S$  &
$E$ \\
\midrule
w/ storage      & 10.5 & 0   & 3.9 & 15.5 \\
w/out storage   & 10.0 & 3.8 & --  & --   \\
\bottomrule
\end{tabularx}
\end{table}

\section{Conclusions}
\label{sec:conclusion}

This paper extends the peak-load pricing model to include storage
that is constrained in both power and stored energy capacity. The resulting
two-period model, together with a set of simplifying assumptions regarding
storage viability, off-peak duration, and the intertemporal value of stored
energy, admits closed-form expressions linking scarcity prices to storage
technology parameters and the length of peak periods.

The on-peak prices required to support storage investment decompose into
a variable-cost component, driven by round-trip efficiency losses, and a
fixed-cost component, driven by storage investment and peak-period duration.
For realistic parameter values, the latter fixed-cost component accounts for
the bulk of the equilibrium scarcity premium. The structure of this fixed-cost
premium differs from that of a conventional generator: the duration constraint
results in energy-capacity costs being recouped on a per-event basis, rather
than being amortized over the annual total of scarcity hours. Our
results reaffirm the importance of considering distinct scarcity events when
evaluating storage, rather than aggregated load durations.

We make strong assumptions to simplify our results as a function of the
on-peak period and the storage parameters, with the resulting benefit of
one-to-one comparison to the known results for peaker plants in
the classic peak-load model. Relaxing these assumptions would introduce
additional terms associated with the off-peak period; exploring these outcomes
remains for future work. However, our underlying takeaway --- duration limits
restrict stored energy capacity costs from being amortized over \textit{all}
scarcity hours --- still holds without adopting these assumptions.

Additionally, the two-period model assumes that storage investment costs are
solely recovered during a peak period. In practice, ``shoulder'' periods of
load can also provide arbitrage opportunities. However, the inelasticity of
electricity demand in conjunction with scarcity prices (e.g. the value of lost load)
which exceed standard energy costs by several orders of magnitude suggest that
even if some profits can be collected during shoulder periods, the majority of
cost recovery will occur during peak periods. Arbitrage
opportunities across multiple periods reduce the level of scarcity prices
needed to support storage investment, but scarcity periods will likely remain
the primary source of cost recovery in equilibrium. Future extensions could
incorporate uncertainty in peak durations or multi-period arbitrage
opportunities to better reflect these real-world conditions
and evaluate the robustness of our claims.

The numerical example illustrates that adding
storage reshapes equilibrium outcomes with a narrowing of price spreads,
substitution for peaker generation, and synergistic investment with baseload
generation.

From a regulatory perspective, our analysis demonstrates that similar to peaker
generators, cost recovery for storage relies on scarcity prices which
dramatically exceed variable costs. If scarcity prices are administratively
capped without accompanying compensation mechanisms (e.g. capacity payments),
storage will be under-incentivized, leading to a suboptimal capacity mix and
lower total system welfare.

\section{Acknowledgment}
ChatGPT was used to assist with grammar and style. The authors would
like to thank the anonymous reviewers for their helpful comments which improved
the paper.

\bibliography{references}

\end{document}